\documentclass[a4paper]{article}

\usepackage{INTERSPEECH2021}
\usepackage{url}
\usepackage[utf8]{inputenc}
\usepackage{graphicx}
\usepackage{amsmath}
\usepackage{booktabs}
\usepackage{algorithm}
\usepackage{algorithmic}
\usepackage{mathtools}
\usepackage{amsfonts}
\usepackage{blindtext}
\usepackage{booktabs}
\usepackage{multirow}
\usepackage{subfigure}

\title{Learning Fine-Grained Cross Modality Excitement for Speech Emotion Recognition}
  
\name{Hang Li, Wenbiao Ding, Zhongqin Wu, Zitao Liu\thanks{The corresponding author: Zitao Liu.}}
\address{TAL Education Group, Beijing, China}
\email{\{lihang4,dingwenbiao,wuzhongqing,liuzitao\}@tal.com}

\begin{document}

\maketitle

\begin{abstract}
  Speech emotion recognition is a challenging task because the emotion expression is complex, multimodal and fine-grained. In this paper, we propose a novel multimodal deep learning approach to perform fine-grained emotion recognition from real-life speeches. We design a temporal alignment mean-max pooling mechanism to capture the subtle and fine-grained emotions implied in every utterance. In addition, we propose a cross modality excitement module to conduct sample-specific adjustment on cross modality embeddings and adaptively recalibrate the corresponding values by its aligned latent features from the other modality. Our proposed model is evaluated on two well-known real-world speech emotion recognition datasets. The results demonstrate that our approach is superior on the prediction tasks for multimodal speech utterances, and it outperforms a wide range of baselines in terms of prediction accuracy. Further more, we conduct detailed ablation studies to show that our temporal alignment mean-max pooling mechanism and cross modality excitement significantly contribute to the promising results. In order to encourage the research reproducibility, we make the code publicly available at \url{https://github.com/tal-ai/FG_CME.git}.
\end{abstract}


\section{Introduction}

Emotion plays an important role in human communication. It has significant effects on information transmission and human interactions. The expression of emotion is usually multimodal, including voice, speech content, facial expressions, etc. With the development of machine learning and deep learning, a large number of speech interaction systems have emerged, such as voice assistants \cite{hoy2018alexa,zhang2016cross,guo2018feature}, intelligent tutoring systems \cite{tekin2015etutor,abdelwahab2015supervised}, etc.  Speech emotion recognition is particularly useful to this kind of systems to better understand the content of the speech and generate a natural response based on the recognized emotion \cite{ostendorf2005human,jin2015speech,arsikere2014computationally}. 

Various types of machine learning methods have been applied to improve the performance of speech emotion recognition. However, the majority of them are not sufficient for accurately detecting speech emotions from human due to the following two challenges. First, \textit{emotion is multimodal}. Both the vocal sound and the linguistic content express emotions. It is insufficient to only consider information from single modality when conducting emotion recognition. For example, sentences expressed in different tones will show different emotions even with the same content. Second, \textit{emotion expression is fine-grained}. The emotion of a sentence is often expressed by specific words or voice fragments. The utterance-level emotion recognition approaches treat every words equally and may fail to capture fine-grained and subtle emotions.

Recently, a large number of approaches have been developed to address above challenges \cite{schuller2004speech,xu2019learning,yoon2018multimodal,huang2019speech}. Researchers have jointly combined multiple modalities for speech emotion recognition and achieved better results compared to methods that only consider information from single modality \cite{schuller2004speech,yoon2018multimodal}. For example, Schuller et al. combined acoustic features and linguistic information in a hybrid architecture to identify emotional key phrases, and assessed the emotional salience of verbal cues from both phoneme sequences and words \cite{schuller2004speech}. Yoon et al. built deep neural networks to learn vocal representations and text representations and concatenated them for emotion classification \cite{yoon2018multimodal}. In spite of the success of above methods, there is still large room for improvement. Most existing approaches are based on utterance-level fusion, such as directly concatenating acoustic and text features. These methods pay little attention to fine-grained multimodal emotional information and are not able to capture subtle emotions in real-life speech conversations. Therefore, it is very necessary to conduct fine-grained learning of spoken speeches and make full use of relationships between acoustic and semantic modalities.

% use low-level multi-modal learning methods. For example, model the two modalities separately, and then fuse the prediction results, or extract raw features of the two modalities separately and then concatenate them to form a new feature vector and then feed it to the model. The limitation of these methods is that there is no deep fusion and interaction between the two modalities. The approach we proposed uses deep neural network and alignment mechanisms to fully interact the information of the two modalities to achieve better results.

% There are many multimodal learning methods that combine the two modalities of speech and text, but most of them 

To overcome above challenges, we propose a novel multimodal neural framework to perform fine-grained emotion recognition. Instead of simply using utterance-level features, we propose a temporal alignment mean-max pooling operation and a cross modality excitation mechanism to capture the interrelations between different modalities. We have conducted extensive experiments on two public available emotion recognition datasets, i.e., Interactive Emotional Dyadic Motion Capture database (IEMOCAP) \footnote{https://sail.usc.edu/iemocap/} and Ryerson Audio-Visual Database of Emotional Speech and Song (RAVDESS) \footnote{https://smartlaboratory.org/ravdess/}. The results show that our method outperforms all baselines.

Overall this paper makes the following contributions:

\begin{itemize}

\item We conduct fine-grained learning with a temporal alignment mean-max pooling operation and cross-modality excitement mechanism to obtain plentiful cross modality information from both voice fragments and sentences.

\item We design and evaluate our approach quantitatively by using two well-known real-world emotion recognition datasets. Besides, we present detailed ablation studies to demonstrate the effectiveness of each component in our proposed model.

\end{itemize}

\section{Related Work}

% speech emotion recognition
Speech emotion recognition has been studied for decades in both the machine learning and speech communities \cite{schuller2003hidden,trigeorgis2016adieu,seehapoch2013speech,lee2011emotion,el2007speech}. Following the mainstream direction, researchers have extracted hand-craft features from audio data and applied them to classic supervised learning methods such as support vector machines \cite{seehapoch2013speech}, hidden Markov models \cite{schuller2003hidden}, Gaussian mixture models \cite{el2007speech}, Tree-Based models \cite{lee2011emotion}, etc. 

With recent advances of deep learning, various neural networks have been used in the task of speech emotion recognition \cite{trigeorgis2016adieu,han2014speech,badshah2017speech,mirsamadi2017automatic}. Han et al. extracted speech signal features from audio and used deep neural networks to classify speech emotion \cite{han2014speech}. Badshah et al. extracted raw spectrograms features and used convolutional neural network (CNN) to extract high-level features to build the model \cite{badshah2017speech}. Due to the sequential structure of audio, Trigeorgis et al. proposed an end-to-end model that employs long short-term memory (LSTM) network to capture the contextual information \cite{trigeorgis2016adieu}. Mirsamadi et al. used recurrent neural networks with local attention mechanism to automatically discover emotionally relevant features from speech and provided more accurate predictions \cite{mirsamadi2017automatic}.

In addition to tackle the speech emotion recognition problem by directly applying deep neural networks on audio datasets, multimodal learning frameworks that jointly consider emotions implied in different modalities \cite{xu2019learning,tzirakis2017end,lee2018convolutional}. For instance, Tzirakis et al. provided a multimodal system to perform an end-to-end spontaneous emotion prediction task from speech and visual data \cite{tzirakis2017end}. Lee et al. proposed an attention model for multimodal emotion recognition from speech and text data, and provided an effective method to learn the correlation between two output feature vectors from separate yet jointly trained CNNs \cite{lee2018convolutional}. 

The closest work to our research is Xu’s \cite{xu2019learning}, where they proposed a fine-grained method to learn alignment between the original speech and the recognized text with attention mechanism. Our approach is different from that: (1) we propose a temporal mean-max alignment pooling method to combine each word and its corresponding voice instead of attention mechanism; and (2) we design a cross modality excitement module to enhance the interactions between the two modalities.

\section{Our Approach}

\begin{figure}[!tpbh]
    \centering
    \includegraphics[width=0.45\textwidth] {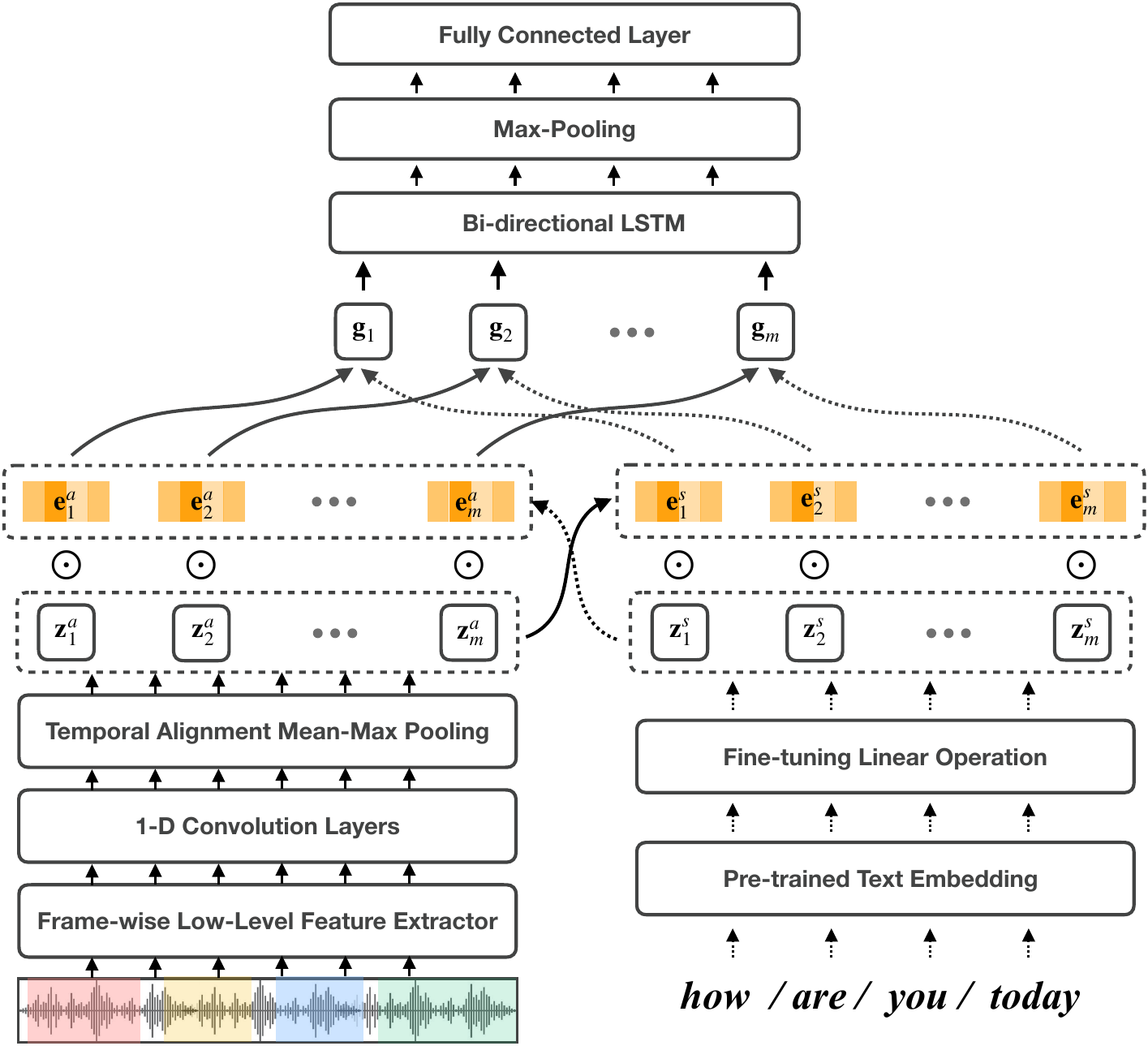}
    \caption{The proposed neural framework.}
    \label{fig:network_structure}
\end{figure}

The architecture of our purposed model is shown in Figure.\ref{fig:network_structure}, which is composed of four key components: (1) unimodal embedding module, which captures the acoustic and semantic information respectively; (2) temporal alignment mean-max pooling, which aggregates the acoustic embedding for each word based on its corresponding speech frames, (3) cross modality excitement module, which converts unimodal information into a fine-grained multimodal representation through the cross modality excitement mechanism; and (4) multimodal prediction, which utilizes a bidirectional LSTM network to model the sequential information within the entire speech sentence. 
% In the following, let $\boldsymbol{a}_i^{(r)}$ be the rhythmic features of the speech signals and $\boldsymbol{s}_i$ be the low-dimensional semantic vectors of speech texts. Let $\boldsymbol{a}^{(e)}_j$ be the fine-grained representation after the CME module.

% \vspace{-0.2cm}
% \vspace{-0.5cm}
 
\subsection{Unimodal Embedding}

\subsubsection{Acoustic Embedding}

For each utterance, we first transform it into $n$ frames $\{f_i\}_{i=1}^n$ of width 25ms and step 10ms. We extract low-level features, i.e., $\mathbf{x}^a_i \in \mathbb{R}^p$, from each frame $f_i$ and the utterance-level feature is $\mathbf{X}^a \in \mathbb{R}^{p \times n}$, $\mathbf{X}^a = [\mathbf{x}^a_1, \cdots, \mathbf{x}^a_n]$. Then we extract the context-aware acoustic embedding $\mathbf{\hat{X}}^a_1 \in \mathbb{R}^{q \times n}$ by feeding $\mathbf{X}^a$ into a multi-layer 1-d CNN, i.e., $\mathbf{\hat{X}}^a_1 = \mathbf{CNN}(\mathbf{X}^a)$. $p$ and $q$ represent the dimensions of low-level features and the extracted acoustic representations respectively.

% Similar to previous studies, we experiment with recurrent neural network architectures, e.g. BiGRU[], BiLSTM[] for encoding the sequential LLFs. 
% And, we find that all these structure receives the comparable results, but CNN structure usually provides us a shorter training and inference time. 

\subsubsection{Semantic Embedding}

For each utterance, we obtain its transcription by an automatic speech recognition (ASR) service. Then, we extract the linguistic representation $\mathbf{X}^s \in \mathbb{R}^{l \times m}$ via pre-trained language models \cite{pennington2014glove}. $m$ denotes the number of words in the ASR transcription. After that, we apply a linear transformation operator $\mathbf{W}^s \in \mathbb{R}^{t \times l}$ on $\mathbf{X}^s$ to obtain the fine-tuned semantic embeddings $\mathbf{Z}^s \in \mathbb{R}^{t \times m}$, i.e., $\mathbf{Z}^s = \mathbf{W}^s \mathbf{X}^s$, which are used in the downstream emotion recognition task. 

\subsection{Temporal Alignment Mean-Max Pooling}

To capture the subtle emotion expression implied in the utterance, we conduct the fine-grained recognition by extracting the underlying word-level acoustic representation. Hence, we propose a temporal alignment mean-max pooling to aggregate the frame-level acoustic representation $\mathbf{\hat{X}}^a$. Instead of the simply averaging all the frame-level acoustic embeddings like mean pooling operator, our temporal alignment pooling operator uses a word-level binary alignment matrix $\mathbf{A}$ to only select relevant frames' acoustic features for each corresponding word. The binary alignment matrix $\mathbf{A} \in \mathbb{R}^{n \times m}$ is a block-wise diagonal matrix, which is calculated by taking an orthographic transcription of an audio file and generating a time-aligned version using a pronunciation dictionary to look up phones for words. The temporal aligned word-level acoustic representation $\mathbf{Z}^a = [\mathbf{z}^a_1,\cdot\cdot\cdot,\mathbf{z}^a_m]$ can be obtained as follows:

% \vspace{-0.2cm}

% $$\mathbf{\hat{z}}^{a}_{j} = (\mathbf{a}_{\cdot,j} \cdot \mathbf{\hat{X}}^{a})\ /\ \sum^N_{i=1} \mathbf{a}_{i,j}$$
$$\mathbf{\hat{z}}^{a}_{j} = \mathrm{meanpool}(\{\mathbf{a}_{\cdot,j}\odot\mathbf{\hat{x}}_{i}^a\ |\ i=1...N\})$$
\vspace{-0.2cm}
$$\mathbf{\tilde{z}}^{a}_{j} = \mathrm{maxpool}(\{\mathbf{a}_{\cdot,j}\odot\mathbf{\hat{x}}_{i}^a\ |\ i=1...N\})$$
\vspace{-0.2cm}
$$\mathbf{z}_j^a = [\mathbf{\hat{z}}^{a}_{j},\mathbf{\tilde{z}}^{a}_{j}]$$

\noindent where $a_{\cdot,j}$ represents the $j$th row of alignment matrix $A$, $\mathbf{z}^a_i$ represents the acoustic feature for $i$th frame. $[\cdot,\cdot]$ denotes the concatenation operation, and $\odot$ represents the element-wise product.

\subsection{Cross Modality Excitation}

% To learn the nonlinear interactions between the acoustic and semantic modalities, we develop a cross modality Adjustment (CMA) module. The CMA module serves as a sample specified adjustment term for the aligned representations from different modalities. Specifically, we add the linear layers to each modality embedding $x^a$, $x^s$, and utilize its projected representation $\mathbf{b}^a$, $\mathbf{b}^s$ as the bias adjustment vector for embeddings from another modalities.

% For semantic embedding $x^s_{i}$, the projected acoustic adjustment term $x^a_{i}$ behaves likes an correction vector which 

% sample specified filter for each dimension value 

% The bias adjustment provides the 

% then we use the projected bias embeddings to the other embedding 

% employ a simple bia adjustment mechanism with a sigmoid activation, i.e., $\mathbf{E} = \delta(\mathbf{W} \mathbf{Z}^s), \mathbf{E} \in \mathbb{R}^{q \times m}$, where $\delta(\cdot)$ denotes the sigmoid function and $\mathbf{W} \in \mathbb{R}^{q \times t}$ represents linear projection operator. $\mathbf{E}$ is the excitation matrix and acts as the semantic weights adapted to the word-specific features. Then the temporal aligned word-level acoustic representation $\mathbf{Z}^a_2$ is adaptively recalibrated by the excitation matrix and the fused multimodal representation $\mathbf{Z}^a$ is computed as follows:

% $$\mathbf{Z}^a = \mathbf{E} \odot \mathbf{Z}^a_2 = \delta(\mathbf{W} \mathbf{Z}^s) \odot \mathbf{Z}^a_2$$

% \noindent where $\odot$ represents the element-wise product. 

To learn the nonlinear interactions between the acoustic and semantic modalities, similar to \cite{hu2018squeeze}, we develop a cross modality excitation (CME) module. The CME module serves as sample-specific activations and is learned for each embedding dimension by a self-gating mechanism based on dimensional dependence. Specifically, we employ a simple gating mechanism with a sigmoid activation, i.e., $\mathbf{E}^a = \delta(\mathbf{W}^a \mathbf{Z}^s), \mathbf{E}^a \in \mathbb{R}^{q \times m}$ and $\mathbf{E}^s = \delta(\mathbf{W}^s \mathbf{Z}^a), \mathbf{E}^s \in \mathbb{R}^{t \times m}$, where $\delta(\cdot)$ denotes the sigmoid function and $\mathbf{W}^a \in \mathbb{R}^{q \times t}, \mathbf{W}^s \in \mathbb{R}^{t \times q}$ represent linear projection operator for the aligned acoustic and semantic representations respectively. $\mathbf{E}^a$ and $\mathbf{E}^s$ are the excitation matrixs, which act as the weights adapted to the corresponding features. Then the temporal aligned word-level acoustic and semantic representations are adaptively calibrated by the corresponding cross modality excitation matrices and the cross modality adapted representations $\mathbf{\bar{Z}}^a$ and $\mathbf{\bar{Z}}^s$ are computed as $ \mathbf{\bar{Z}}^a = \mathbf{E}^a \odot \mathbf{Z}^a$ and $\mathbf{\bar{Z}}^s = \mathbf{E}^s \odot \mathbf{Z}^s$.

% $$\mathbf{\bar{Z}}^a = \mathbf{E}^a \odot \mathbf{Z}^a = \delta(\mathbf{W} \mathbf{Z}^s) \odot \mathbf{Z}^a$$
% \vspace{-0.2cm}
% $$\mathbf{\bar{Z}}^s = \mathbf{E}^s \odot \mathbf{Z}^s = \delta(\mathbf{W} \mathbf{Z}^a) \odot \mathbf{Z}^s$$

% \noindent where $\odot$ represents the element-wise product. 

\subsection{Multimodal Prediction}

Let $\mathbf{\bar{z}}_i^a$ and $\mathbf{\bar{z}}_i^s$ be the acoustic and semantic representations of \emph{i}th word in the original utterance, i.e., $\mathbf{\bar{Z}}^a = \{\mathbf{\bar{z}}_i^a\}_{i = 1}^m$ and $\mathbf{\bar{Z}}^s = \{\mathbf{\bar{z}}_i^s\}_{i = 1}^m$. Each word's multimodal representation $\mathbf{g}_i$ is computed by concatenating $\mathbf{\bar{z}}_i^a$ and $\mathbf{\bar{z}}_i^s$, i.e., $\mathbf{g}_i = [\mathbf{\bar{z}}_i^a, \mathbf{\bar{z}}_i^s]$. In the multimodal prediction layer, we utilize a bi-directional LSTM to capture the word-level sequential patterns on top of the multimodal representation of each word $\mathbf{g}_i$. The resulting hidden state of $\mathbf{h}_i$ is the concatenation of hidden representations from both directions, i.e., $\mathbf{h}_i = [\overrightarrow{\mathbf{h}_i},\overleftarrow{\mathbf{h}_i}]$. After that, a max-pooling layer is applied to aggregate the sequential information for all the hidden states $\mathbf{h}_i$s in the utterance sequence, i.e., $\mathbf{h}_i^* = \mathrm{maxpool}(\{\mathbf{h}_i\}_{i=1}^m)$. Finally, we use a two-layer fully-connected feed forward network (FCN) to conduct the final predictions, i.e., 

\begin{equation}
    \mathbf{p}_i = \mathrm{softmax}(\mathrm{FCN}(\mathbf{h}_i^*)) 
\end{equation}

\noindent where $\mathbf{p}_i$ is the probabilistic vector that indicates the final probabilities of class memberships of utterance $i$. In this work, we use the multi-class cross-entropy loss to optimize the prediction accuracy, which is defined as follows:

\begin{equation}
    \mathcal{L}=-\sum_{i=1}^{N}\sum_{k=1}^{K}\ y_{i,k}\log p_{i,k}
\end{equation}

\noindent where $p_{i,k}$ is the $k$th element of $\mathbf{p}_i$, and $y_{i,k} = 1$ if the $i$th sample belongs to the $k$th class.

\section{Experiments}

\begin{figure*}[!hptb]
    \centering
    \subfigure[]{\includegraphics[width=0.23\textwidth]{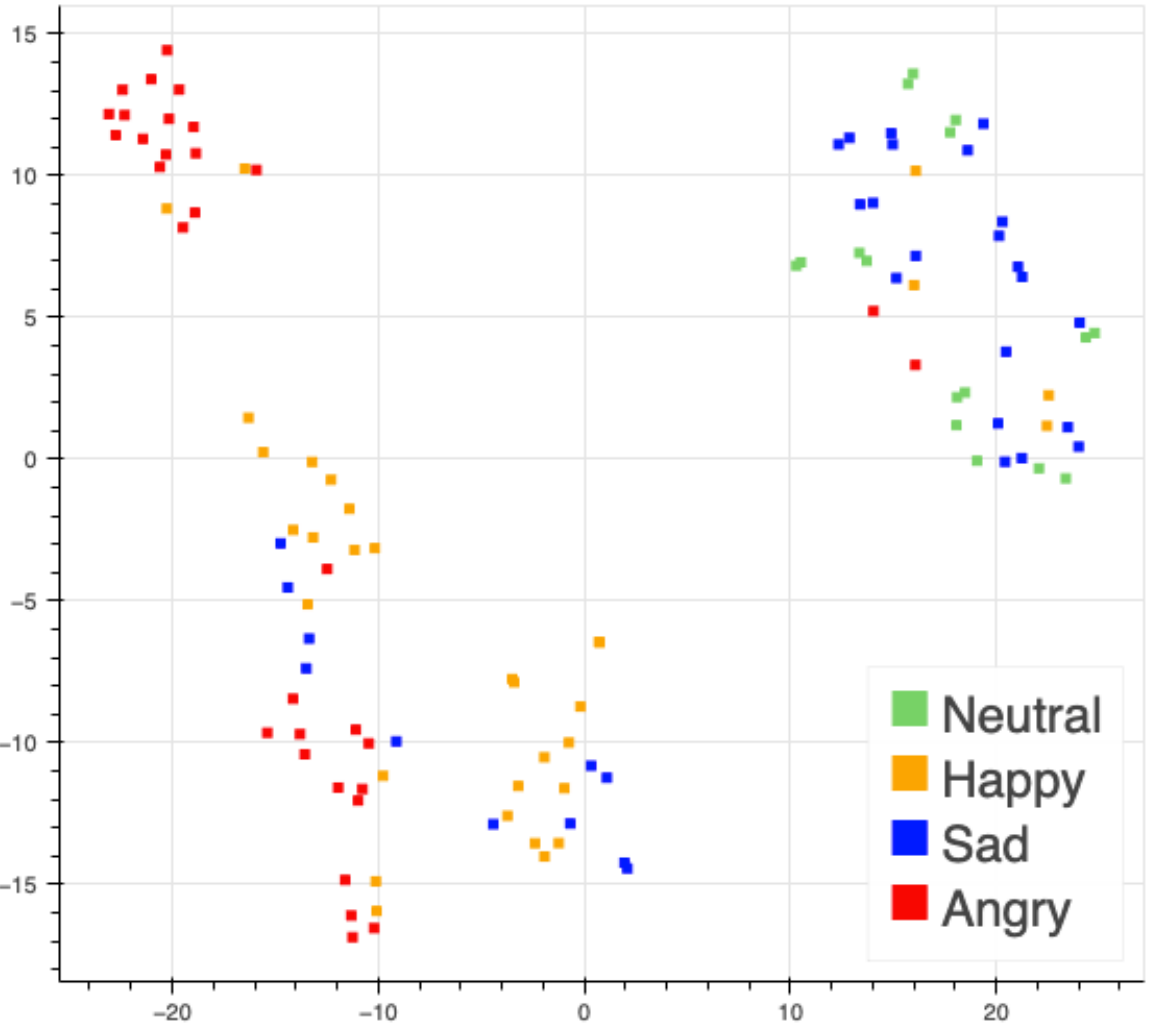}} \hfill
    \subfigure[]{\includegraphics[width=0.23\textwidth]{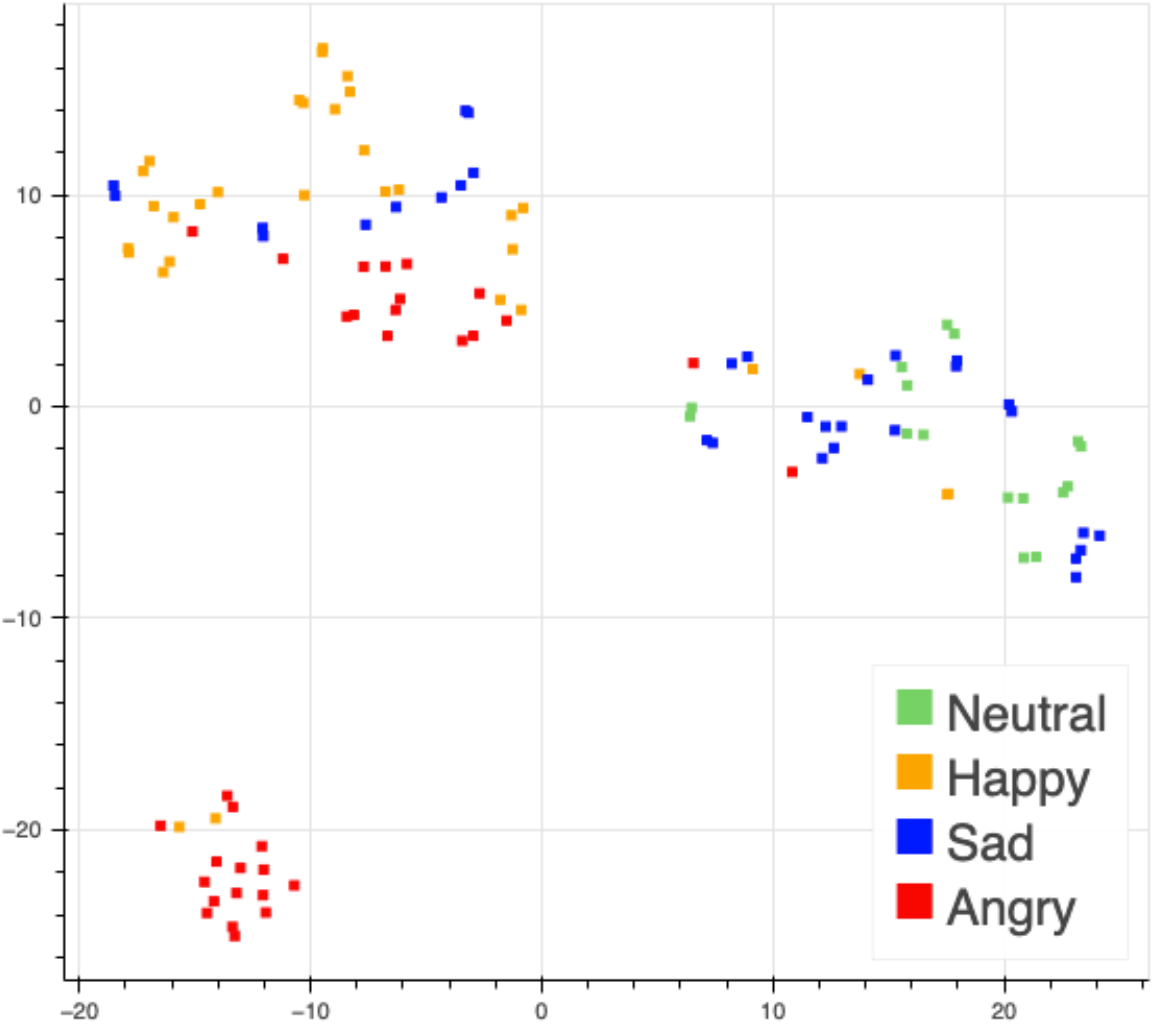}} \hfill
    \subfigure[]{\includegraphics[width=0.23\textwidth]{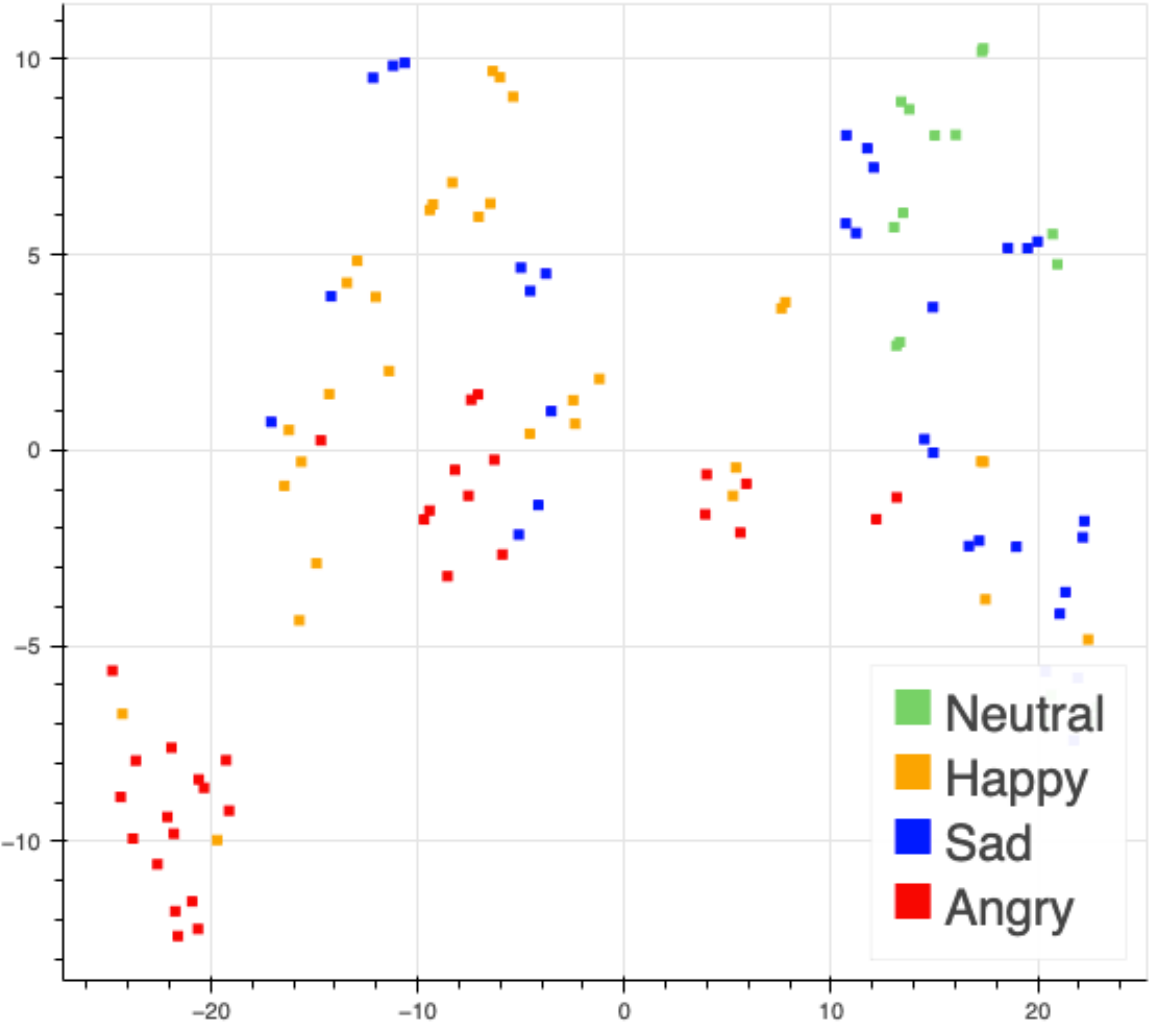}} \hfill
    \subfigure[]{\includegraphics[width=0.23\textwidth]{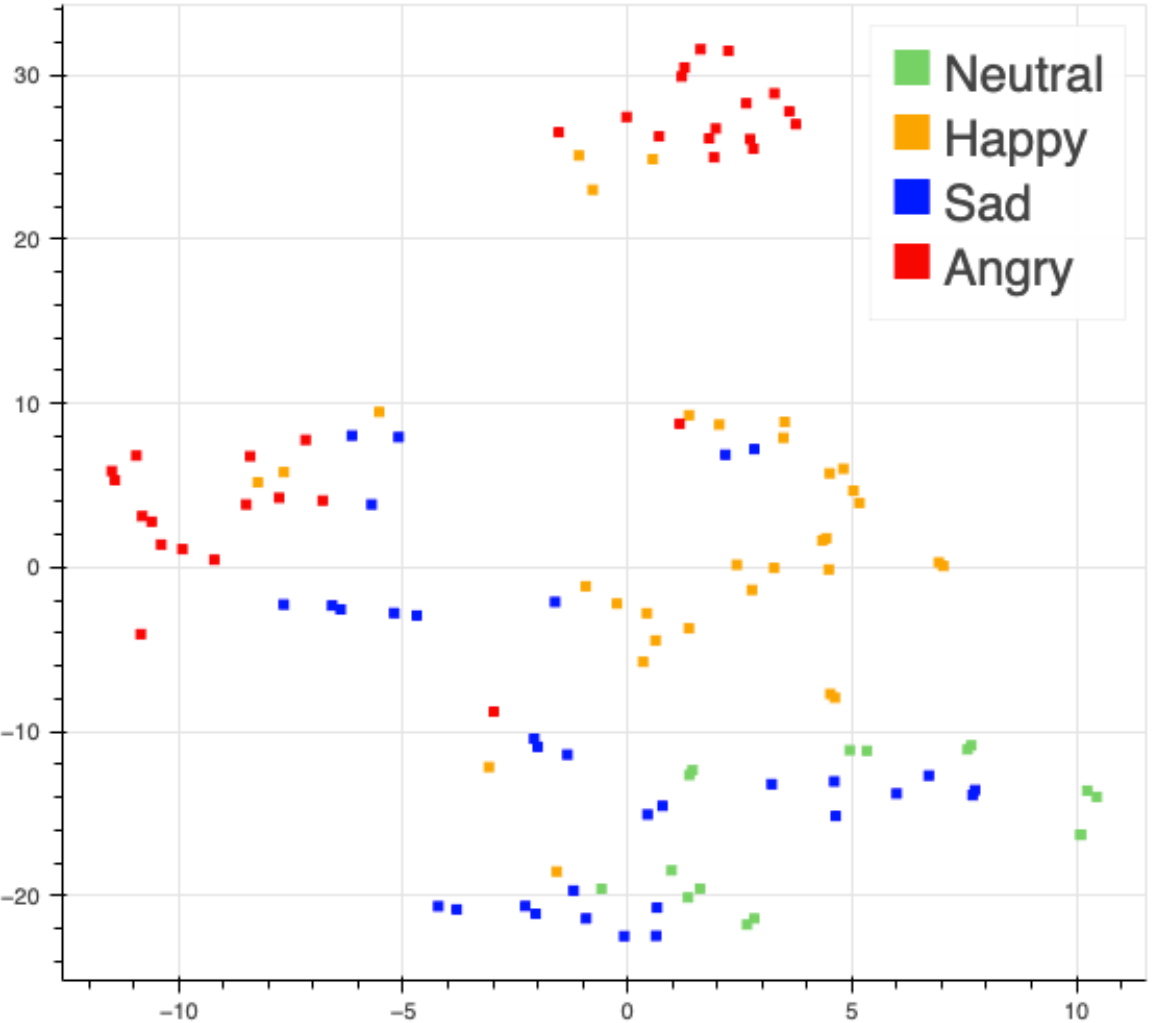}}
    \vspace{-0.2cm}
    \caption{Visualization of embeddings for validation samples from RAVDESS with different ablated components. (a) w/o both excitement (b) w/o semantic excitement (c) w/o acoustic excitement (d) with both excitement}
    \label{fig:visual}
\end{figure*}
\vspace{-0.2cm}

\subsection{Data}
To assess our proposed framework, we conduct several experiments with two open source English emotion datasets: Interactive Emotional Dyadic Motion Capture database (IEMOCAP) \cite{Busso2008IEMOCAPIE} \footnote{\url{https://sail.usc.edu/iemocap/}} and Ryerson Audio-Visual Database of Emotional Speech and Song (RAVDESS) \cite{livingstone} \footnote{\url{https://smartlaboratory.org/ravdess/}}. 

\subsubsection{IEMOCAP} 

IEMOCAP consists of five sessions of dyadic conversations between different pairs of actors. The total number of utterances in IEMOCAP is 10,039. For each utterance, the audio, transcriptions, video and actor's motion-capturing recordings are collected and emotion labels are annotated by the majority voting results from three experienced evaluators. Following the similar experiment settings with prior studies \cite{xu2019learning,yoon2018multimodal}, we use four emotions (\emph{angry}, \emph{happy}, \emph{neutral} and \emph{sad}) for classification evaluation. Besides that, we perform the five-fold cross validation and the average results are reported.  

\subsubsection{RAVDESS} 

RAVDESS is a validated multimodal database of emotional speeches and songs, which includes 9 hours of speeches and 1,440 utterance. Unlike the IEMOCAP, the speech content of each utterance in RAVDESS is the same and actors express the different emotions through their different tones. The label of each utterance is validated by multiple raters and the intensities of emotions are conducted based on raters' responses on emotion strength. In our experiment, we use the same four emotions (\emph{angry}, \emph{happy}, \emph{neutral} and \emph{sad}) for classification. The five-fold cross validation is also incorporated for the robustness of the final conclusions.

\subsection{Baselines}
To evaluate the effectiveness of our proposed framework, we carefully choose the following state-of-the-art emotion classification approaches  as our baselines: 
(1) \textbf{Acoustic-CNN+LSTM}($\mathrm{CNN+LSTM}^{a}$): We train the similar model proposed by Satt et al., which uses CNN for extracting acoustic embeddings and applies the LSTM layer at the top for aggregating utterance embedding \cite{satt2017efficient}.
(2) \textbf{Acoustic-LSTM+Attention}($\mathrm{LSTM+Attn}^{a}$): A bi-directional LSTM is used for generating the acoustic embeddings from the frame-wise low-level features. An attention layer is used for the final aggregation \cite{mirsamadi2017automatic}.
(4) \textbf{Semantic-LSTM+Attention}($\mathrm{LSTM+Attn}^{s}$): Similar to $\mathrm{LSTM+Att}^{a}$) but instead only the text features are used.
(5) \textbf{Multimodal-Utterance+Concat}($\mathrm{UttConcat}^{m}$): Two individual bi-directional LSTMs are used as unimodal encoders for acoustic and semantic features and the unimodal embeddings are concatenated at utterance level for the final prediction \cite{yoon2018multimodal}.
(6) \textbf{Multimodal-AttentionFusion}($\mathrm{AttnFusion}^{m}$): Similar to $\mathrm{Utt+Concat}^{m}$ but instead applying the multi-head attention for fusing the multimodal features at word level \cite{xu2019learning}.
(7) \textbf{Multimodal-MultihopAttention}($\mathrm{MHA}^{m}$): Similar to $\mathrm{UttConcat}^{m}$ and $\mathrm{AttnFusion}^{m}$, except that a multi-hop attention mechanism is used for the inference of correlations between two modalities \cite{yoon2019speech}.

\subsection{Implementation Details}
For the acoustic embeddings, we first transform sample's audio signal into frames with 25ms and step size 10ms. Then, we use a Python Library \cite{Gian2015pyAudioAnalysis} to extract the 40-dimensional filterbank features from each frame. The 1d CNN network we used for embedding the acoustic features has 3 layers with kernel sizes (5, 2, 2) and the stride size of each kernel is set to 1 for keeping the length of the frame sequences. The number of the filters are set to (64,128,128) and the dimension of the output acoustic embedding $\mathbf{Z}^a$ is 256, which is generated by concatenating the corresponding word level mean-pooling and max-pooling embeddings.

For the semantic embeddings, we use a 300-dimensional pre-trained GloVe embedding \cite{pennington2014glove} for mapping the words into the fixed-length vectors for each utterance transcription. The dimension of  linear operator $\mathbf{W}^s$ we used for fine-tuning the pre-trained word embeddings is 256. To implement the model, we use 256 hidden units in bidirectional LSTM, i.e. $d_{h_i}$, in the multimodal sequence feature extracting layer and the neurons of the two-layer full-connected layer $\mathrm{FCN}(\cdot)$ is (128,4). To train our model, we use Adam optimizer \cite{kingma2014adam} with learning rate of 0.0005.

\subsection{Results \& Analysis} 

We evaluate and compare the performance of different method based on two widely used metrics for emotion classification: weighted accuracy (WA) that is the overall classification accuracy and unweighted accuracy (UA) that is the average recall over the emotion categories.

\begin{table}[!hptb]
    \centering
    \caption{Experimental results on \emph{IEMOCAP} and \emph{RAVDESS} datasets. $\mathrm{UA}$ and $\mathrm{WA}$ indicate the unweighted accuracy and weighted accuracy}
    % \begin{small}
    \begin{tabular}{@{}lccccc@{}} \toprule
    \label{tab:results}
        & \multicolumn{2}{c}{\textbf{IEMOCAP}} & \multicolumn{2}{c}{\textbf{RAVDESS}} \\
        & $\mathrm{WA}$ & $\mathrm{UA}$ & $\mathrm{WA}$ & $\mathrm{UA}$ \\ \hline
        \bf{Acoustic only}                       & & & & \\
        $\mathrm{CNN+LSTM}^{a}$        & 0.635 & 0.588 & 0.640 & 0.611  \\ 
        $\mathrm{LSTM+Attn}^{a}$        & 0.573 & 0.574 & 0.665 & 0.627  \\ \hline
        \bf{Semantic only}                       & & & & \\
        $\mathrm{LSTM+Attn}^{s}$        & 0.640 & 0.647 & $-$ & $-$  \\ \hline
        \bf{Semantic + Acoustic}            & & & & \\
        $\mathrm{UttConcat}^{m}$       & 0.700 & 0.677 & 0.629 & 0.615  \\
        $\mathrm{AttnFusion}^{m}$       & 0.712 & 0.725 & 0.634 & 0.593  \\
        $\mathrm{MHA}^{m}$       & 0.678 & 0.688 & 0.615 & 0.572  \\\hline
        \bf{Ablation Studies}            & & & & \\
        w/o $\mathrm{Excite}^{a}$ & 0.713 & 0.722 & 0.703 & 0.683 \\ 
        w/o $\mathrm{Excite}^{s}$ & 0.721 & 0.732 & \bf{0.720} & 0.696 \\ 
        w/o $\mathrm{Excite}^{m}$ & 0.712 & 0.720 & 0.718 & 0.700 \\ 
        w/ CTC Align & 0.657 & 0.688 & 0.582 & 0.566 \\ \hline
        Ours & \bf{0.727} & \bf{0.735} & \bf{0.720} & \bf{0.708} \\ \bottomrule
    \end{tabular}
    % \end{small}
\end{table}

The results of our experiments show that our approach outperforms all other methods on both IEMOCAP and RAVDESS. Specifically, from Table.\ref{tab:results}, we find the following results: (1) When comparing the unimodal based models ($\mathrm{CNN+LSTM}^{a}$, $\mathrm{LSTM+Attn}^{a}$,$\mathrm{LSTM+Attn}^{s}$) with the multimodal ones ($\mathrm{UttConcat}^{m}$, $\mathrm{AttnFusion}^{m}$, $\mathrm{MHA}^{m}$), we find that the multimodal features provide a huge boost for the models' performance on both WA and UA. (2) Comparing $\mathrm{UttConcat}^{m}$ and $\mathrm{MHA}^{m}$ with $\mathrm{AttnFusion}^{m}$, the fusion of multimodal features in a fine-grained manner provides the model with a strong capability to capture the subtle emotions expressed within the utterance level.

\subsection{Ablation Studies}

Apart from applying the comparisons above, we also present the ablation results of different key components in our proposed algorithm in Table.\ref{tab:results}. Overall, each component of our proposed model plays an important role in improving the model's performance on both datasets. For model without acoustic excitement (w/o $\mathrm{Excite}^a$), we find its performance drops dramatically in RAVDESS, which is consistent with our expectations since samples in the RAVDESS share the similar text inputs. Comparing model without any excitement (w/o $\mathrm{Excite}^m$) with model without either acoustic excitement or semantic excitement (w/o $\mathrm{Excite}^s$), we observe that the model's performance is boosted by both excitements and the best performance is achieved by using both excitement modules simultaneously. At last, we replace our pre-calculated word-level binary alignment matrix $A$ with the CTC predicting alignment matrix (w/ CTC Align) \cite{tsai2019multimodal}, we find the model's performance suffers a great loss, which indicates that the pre-calculated alignment matrix $A$ is more compatible with our proposed algorithm.

To provide a better view about the effectiveness of each components, we use t-SNE to visualize the embeddings of validation samples from RAVDESS generated by different models in Fig.\ref{fig:visual}. From the figure, we observe that embeddings of different emotion category's samples from the model without both excitement modules is hard to be separated. After employing the CME block, the intra-distance between different categories is enlarged, which makes it easy for the final classifier to generate a robustness split boundaries for each class.

\section{Conclusion}

In this paper, we propose a novel multimodal deep learning framework to perform fine-grained learning of voice and text in speeches. 
Through applying the temporal alignment operator combining with the cross modality excitation module, we successfully generated a powerful multimodal representation for each utterance in a fine-grained manner. 
The experiments on two open source English emotion datasets: IEMOCAP and RAVDESS demonstrates the effectiveness of our purposed algorithm. Besides, the influence of each component of our model is presented via the detailed ablation studies and analysis. 

\section{Acknowledgements}
This work was supported in part by National Key R\&D Program of China, under Grant No. 2020AAA0104500 and in part by Beijing Nova Program (Z201100006820068) from Beijing Municipal Science \& Technology Commission.

\bibliographystyle{IEEEtran}

\bibliography{icassp2021}

% \bibliography{mybib}

% \begin{thebibliography}{9}
% \bibitem[1]{Davis80-COP}
%   S.\ B.\ Davis and P.\ Mermelstein,
%   ``Comparison of parametric representation for monosyllabic word recognition in continuously spoken sentences,''
%   \textit{IEEE Transactions on Acoustics, Speech and Signal Processing}, vol.~28, no.~4, pp.~357--366, 1980.
% \bibitem[2]{Rabiner89-ATO}
%   L.\ R.\ Rabiner,
%   ``A tutorial on hidden Markov models and selected applications in speech recognition,''
%   \textit{Proceedings of the IEEE}, vol.~77, no.~2, pp.~257-286, 1989.
% \bibitem[3]{Hastie09-TEO}
%   T.\ Hastie, R.\ Tibshirani, and J.\ Friedman,
%   \textit{The Elements of Statistical Learning -- Data Mining, Inference, and Prediction}.
%   New York: Springer, 2009.
% \bibitem[4]{YourName17-XXX}
%   F.\ Lastname1, F.\ Lastname2, and F.\ Lastname3,
%   ``Title of your INTERSPEECH 2021 publication,''
%   in \textit{Interspeech 2021 -- 20\textsuperscript{th} Annual Conference of the International Speech Communication Association, September 15-19, Graz, Austria, Proceedings, Proceedings}, 2020, pp.~100--104.
% \end{thebibliography}

\end{document}